\documentclass[%
 reprint,
 amsmath,amssymb,
 aps,
]{revtex4-1}

\usepackage{graphicx}
\usepackage{dcolumn}
\usepackage{bm}

\usepackage{lipsum}
\usepackage{url}
\usepackage{hyperref}
\usepackage{float}
\usepackage{adjustbox}
\usepackage{multirow}

\makeatletter
\newcommand*{\rom}[1]{\expandafter\@slowromancap\romannumeral #1@}
\makeatother

\begin{document}

\preprint{APS/123-QED}

\title{Existence of Chandrasekhar's limit in GUP white dwarfs}

\author{Arun Mathew}
\email{a.mathew@iitg.ac.in}
\author{Malay K. Nandy}%
 \email{mknandy@iitg.ac.in}
\affiliation{Department of Physics, Indian Institute of Technology Guwahati, Guwahati 781039, India
}%

\date{January 5, 2020}

\begin{abstract}

Various recent theoretical investigations suggest that gravitational collapse of white dwarfs is withheld for arbitrarily high masses if the equation of state is described by the generalized uncertainty principle (GUP). There have been a few attempts to restore the Chandrasekhar limit but they are found to be inadequate from different perspectives and some of them led to unphysical mass-radius relations. In this paper, we rigorously resolve this problem by  analyzing the dynamical instability in general relativity. We confirm the existence of Chandrasekhar's limit as well as stable mass-radius curves that behave consistently with astronomical observations. Moreover, this stability analysis suggests gravitational collapse beyond the Chandrasekhar limit signifying the possibility of compact objects denser than white dwarfs. 

\end{abstract}

\maketitle


\section{Introduction} 

It has recently been argued that the generalized uncertainty principle (GUP) removes the Chandrasekhar limit \cite{Rashidi2016,Moussa2015,Ong2018b,Ong2018a,Mathew2018a}. This is due to the fact that the inclusion of GUP,
 \begin{equation}
 \Delta x \Delta p \geq \frac{\hbar}{2}\left\{ 1+\beta (\Delta p)^2  \right\},
 \end{equation}
via the equation of state gives white dwarfs of excessively high masses. In other words, the mass is no longer bound from above, so that 
\begin{equation}
M_{\rm GUP} = 12\sqrt{2} \left( \frac{\hbar c}{G}\right)^{3/2}\frac{1}{(\mu_e m_u)^2} \ (\beta p_{\rm Fc}^2)^{3/4}.
\end{equation}
This implies that GUP-enhanced equation of state prevents gravitational collapse and halts the formation of compact astrophysical objects denser than white dwarfs. This prediction contradicts astronomical observations that confirm the existence of pulsars \cite{Hewish1968,Mignani2007,Cromartie2019} and black holes \cite{Burke2012,Walsh2013,Liu2019}.  Moreover, it has been observed that the masses of (non-magnetic) white dwarfs fall well within the Chandrasekhar limit \cite{Shipman1972,Shipman1979,Kilic2007},
\begin{equation}
M_{\rm Ch} = 2.0182 \frac{\sqrt{3\pi}}{2} \left( \frac{\hbar c}{G}\right)^{3/2} \frac{1}{(\mu_e m_u)^2} \approx \frac{5.76}{\mu_e^2} M_{\odot}, 
\end{equation}
apart from the super-Chandrasekhar white dwarfs, that may well be double-degenerate mergers \cite{Hicken2007,Silverman2011,Kerkwijk2013}. 

A solution to the problem was proposed by imposing a cutoff in the Fermi momentum at the netronization threshold \cite{Mathew2018a}. Since the process of neutronization is not built into the dynamical equations, and it is imposed {\it by hand}, this solution is not a dynamical consequence of the theory. A more satisfying solution ought to be based on a theory where a collapse happens as a dynamical consequence of the underlying equations of the theory. 

It is important to note that excessive mass of white dwarfs results from the fact that the GUP parameter $\beta$ is positive. In an attempt to resolve this problem, Ong \cite{Ong2018b} assumed the GUP parameter $\beta$ to be negative and obtained a mass-radius relation heuristically from the GUP as
\begin{equation}
R =  \sqrt{|\beta|} \frac{Mm_e^{1/3}c}{\sqrt{M_{\rm Ch}^{2/3}-M^{2/3}}}  \ell_P,
\end{equation}
in the relativistic limit, giving the Chandrasekhar mass $M_{\rm Ch}$ as an upper bound. However, this mass-radius relation has two issues: (i) as the mass $M$ increases, the radius $R$ also increases and (ii) the radius diverges as the mass approaches the Chandrasekhar limit, preventing the formation of compact objects as the density would have been infinitely diluted. These issues are in contradiction with observations that indicate that the radius decreases with increase in mass of white dwarf. Moreover we expect the formation of highly dense objects such as neutron stars or black holes when the mass exceeds the Chandrasekhar limit.

Apart from these issues, there is a more fundamental issue related to the sign of the GUP parameter $\beta$. Theories of quantum gravity suggest a grainy structure of the spacetime which naturally implies a minimum uncertainty in position measurement \cite{Padmanabhan1985a,Padmanabhan1985b,Padmanabhan1986,Padmanabhan1987,Greensite1991}. The minimum uncertainty in length in the GUP scenario is given by $\Delta x_{\min} = \hbar \sqrt{\beta}\sqrt{1+\beta\langle \bf{p} \rangle^2}$ as shown in Ref. \cite{Kempf1995}. This obviously means that $\beta$ must be positive as $\Delta x_{\min}$ is a real-valued quantity. Thus the assumption of $\beta$ being negative is untenable. The positivity of the GUP parameter $\beta$ is also apparent in the thought experiment of observing an electron through a Heisenberg microscope \cite{Mead1964}.  The additional uncertainty in position of the electron due to gravitational interaction with the photon turns out to be a positive quantity, of the order of $\ell_P^2 \Delta p/\hbar$ \cite{Adler1999}, implying $\beta$ is positive. Moreover, a string theoretic consideration with a length scale $\ell_*$ also leads to the same additional uncertainty in position with  $\ell_*$ replacing $\ell_P$ \cite{Amati1989,Konishi1990}. In addition, measurement of the radius of an extremal black hole by dropping a photon into it and by observing the remitted photon gives a similar positive estimate for the uncertainty \cite{Maggiore1993,Maggiore1994,Garay1995}.

Ong and Yao \cite{Ong2018a} suggested an alternative approach to circumvent the problem of non-existence of the Chandrasekhar mass via extending the GUP by incorporating the effect of cosmological constant $\Lambda$, so that 
 \begin{equation}
 \Delta x \Delta p \geq \frac{\hbar}{2}\left\{ 1+\beta (\Delta p)^2 - \lambda \frac{(\Delta x)^2}{L_{\Lambda}^2} \right\}, 
 \end{equation}
with $L_{\Lambda}^2 = \lambda/\Lambda$ which is positive for de-Sitter expansion of the Universe ($\lambda=+3$). Although the observed value of $\Lambda$ is very small, namely $\Lambda\sim10^{-52}$ m$^{-2}$, they showed that this reformulation of GUP leads to a mass-radius relation whose physically acceptable solution is strongly dominated by the cosmological terms and the contribution from $\beta$ is insignificant, making the sign of $\beta$ irrelevant. This mass-radius relation clearly shows that the Chandrasekhar mass is the upper bound. However, this mass-radius relation also suffers from the same two issues as described earlier, namely (i) and (ii) above. Consequently, this reformulation does not represent realistic white dwarfs for which the radius decreases with increasing mass. 

The above discussion suggests that we need not consider the effect of the expanding Universe on white dwarfs. On the other hand, if we include the effect of quantum gravity via the GUP, we must take $\beta$ to be positive, as discussed previously. However, this poses a well-known problem that the Chandrasekhar limit ceases to exist. Since white dwarfs are found below the Chandrasekar limit, it is extremely important to solve this problem posed by GUP. 
A satisfactory model of white dwarfs ought to be based on a rigorous treatment of the gravitational field so that the gravitational collapse for a sufficiently massive white dwarf is well-represented.

In this paper, we present a complete and rigorous approach to resolve this problem. We take the framework of general relativity (GR) and calculate the stellar structure of white dwarfs for positive GUP parameter $\beta$.  We also carry out a dynamical stability analysis of the equilibrium configurations so that the maximal stable configuration is identified. In this way we confirm that the Chandrasekar limit reappears when we take the GUP parameter $\beta$ within the upper bound suggested by the electroweak limit \cite{Das2008}. In fact we find that the Chandrasekar limit robustly exists even when the value of $\beta$ is higher than the electroweak bound. However, this enhancement of $\beta$ has an upper limit $\bar{\beta}$ for the existence of the Chandrasekar mass.

The reminder of the paper is organized as follows. In Section \ref{FEOS} we present the fermionic equation of state following from GUP. In Section \ref{MR} we give details of the mass-radius relation in the framework of general relativity. Section \ref{Stability} presents  the dynamical stability analysis confirming the Chandrasekar limit. A discussion and conclusion is given in Section \ref{Conclusion}.

\vspace{0.5cm}

\section{Generalized uncertainty principle and fermionic equation of state}\label{FEOS} 

A minimum uncertainty in length  due to the granular structure of space, which is essentially a quantum gravitational effect, can be incorporated by generalizing the Heisenberg commutation relations \cite{Kempf1995} to 
\begin{eqnarray}
&&[\hat{x}_i,\hat{p}_j]= i\hbar \delta_{ij}(1+\beta \hat{\textbf{p}}^2), \nonumber \\
&&[\hat{p}_i,\hat{p}_j]=0, \\ 
&&[\hat{x}_i,\hat{x}_j] = 2i\hbar \beta ( \hat{p}_i\hat{x}_j-\hat{p}_j\hat{x}_i ). \nonumber
\end{eqnarray}

These generalized commutation relations incorporate a modified high momentum behavior via the terms containing $\beta\sim\hbar^2/\ell_P^2$, where $\ell_P=\sqrt{G/\hbar c^3}=1.6162\times10^{-33}$ cm is the Planck length. Considering a classical Liouville's equation, it was shown \cite{Chang2002} that the invariant measure of the phase volume take up a factor of $(1+\beta {\bf p}^2 )^{-3}$. This imposes a severe restriction on the allowed quantum states and thus modifies the thermodynamic properties with respect to the ideal case.

The inclusion of quantum gravitational fluctuations via the generalized uncertainty principle in the equation of state of a degenerate electron gas was studied earlier in different contexts \cite{Wang2010,Zhang2014,Moussa2014,Mathew2018a,Nozari2006}. In this section, we present the number density $n$, pressure $P$ and energy density $\varepsilon$ of the electron degenerate gas. With the modified phase volume, we employ the standard method of statistical mechanics to the relativistic electron gas assuming $T=0$, yeilding
\begin{equation} \label{}
n = \frac{8\pi}{h^3}\int_0^{p_F} \frac{p^2 dp}{(1+\beta p^2)^3},
\end{equation}
and
\begin{equation} \label{}
P = \frac{8\pi}{h^3}\int_0^{p_F} \frac{p^2 dp}{(1+\beta p^2)^3} (E_F-E_{\bf p}),
\end{equation}
leading to 
\begin{equation} \label{pressure_1}
n(\xi)= \frac{K}{m_e c^2} \tilde{n}(\xi), \hspace{0.3cm} {\rm and }   \hspace{0.3cm}   P(\xi) = K \tilde{P}(\xi),  
\end{equation}
where
\begin{equation}\label{n}
 \tilde{n}(\xi) = \frac{1}{\alpha^3} \left[ \tan^{-1} (\alpha \xi) -\frac{\alpha \xi(1-\alpha^2\xi^2)}{(1+\alpha^2\xi^2)^2}\right]
\end{equation}
and
\begin{widetext}
\begin{equation}\label{pressure}
\tilde{P}(\xi) =    \frac{\sqrt{1+\xi^2}}{\alpha^3} \left\{  \tan^{-1} (\alpha \xi)   
-\frac{\alpha \xi}{(1-\alpha^2)(1+\alpha^2\xi^2)} \right\} +\frac{1}{(1-\alpha^2)^{3/2}}  \tanh^{-1}\frac{\xi\sqrt{1-\alpha^2}}{\sqrt{1+ \xi^2}}
\end{equation}
\end{widetext}
with $\xi=p_F/m_ec$, $p_F$ being the Fermi momentum, $\alpha=\beta m_e^2 c^2=\beta_0 m_e^2/M_P^2$ ($M_P=\sqrt{\hbar c/G}=2.1765\times10^{-5}$ g) and $K=\pi m_e^4 c^5/h^3$.

The internal kinetic energy $\varepsilon_{\rm int}(\xi)$ of the electron gas for $T=0$ is given by 
\begin{equation}\label{}
\varepsilon_{\rm int} (\xi) = \frac{8\pi}{h^3}\int_0^{p_F} \frac{p^2 dp}{(1+\beta p^2)^3} \left\{ \sqrt{p^2c^2+m_e^2c^4} -m_ec^2\right\}.
\end{equation}

In the dimensionless quantities, the above equation becomes 
\begin{equation}\label{}
\varepsilon_{\rm int} (\xi) = \frac{8\pi m_e^4 c^5}{h^3}  \int_0^\xi \frac{\xi'^2 d\xi'}{(1+\alpha^2 \xi'^2)^3} \left\{ \sqrt{\xi'^2+1} -1\right\},
\end{equation}
leading to 
\begin{widetext}
\begin{equation}\label{}
\varepsilon_{\rm int} (\xi) =   \left\{ \frac{\xi\sqrt{1+\xi^2}[1+(2-\alpha^2)\xi^2]}{(1-\alpha^2)(1+\alpha^2 \xi^2)^2} - \frac{1}{(1-\alpha^2)^{3/2}} \tanh^{-1} \frac{\xi \sqrt{1-\alpha^2}}{\sqrt{1+\xi^2}} \right\}  -\tilde{n}.
\end{equation} 
\end{widetext}
The rest mass density $\rho_0(\xi)= m_u \mu_e n(\xi)$ is related to the energy density as $\varepsilon(\xi) = \rho_0 (\xi) c^2 +\varepsilon_{\rm int}(\xi)$,  where $m_u=1.6605\times10^{-24}$ g is the atomic mass unit and $\mu_e = A/Z$, with $A$ the mass number and $Z$ the atomic number. 
Thus the energy density 
\begin{equation}\label{energydensity_1}
\varepsilon(\xi) = \frac{K}{q} \tilde{\varepsilon}(\xi),
\end{equation} 
where $q = m_e/\mu_e m_u$ and the dimensionless energy density $\tilde{\varepsilon}(\xi)$ is given by 
\begin{widetext}
\begin{equation}\label{energydensity}
\tilde{\varepsilon}(\xi) =  (1-q)\tilde{n} + q \left\{ \frac{\xi\sqrt{1+\xi^2}[1+(2-\alpha^2)\xi^2]}{(1-\alpha^2)(1+\alpha^2 \xi^2)^2} \right. 
\left. - \frac{1}{(1-\alpha^2)^{3/2}} \tanh^{-1} \frac{\xi \sqrt{1-\alpha^2}}{\sqrt{1+\xi^2}} \right\},
\end{equation} 
\end{widetext}

In the high Fermi momentum limit,  that is $\xi\rightarrow\infty$, 

\begin{equation}
\tilde{n}(\xi) \longrightarrow \frac{\pi}{2\alpha^3} = k_1,
\end{equation}
\begin{equation}
 \tilde{P}(\xi) \longrightarrow k_1 \xi  - k_2
\end{equation}
and 
\begin{equation}
\tilde{\varepsilon} \longrightarrow k_1 (1-q)  + q k_2 = 3\kappa
\end{equation}
with
\begin{equation}
k_2 = \frac{1}{\alpha^4} \frac{(2-\alpha^2)}{(1-\alpha^2)} -\frac{\tanh^{-1} \sqrt{1-\alpha^2}}{(1-\alpha^2)^{3/2}}.
\end{equation}
where $k_1$, $k_2$ and $\kappa$ are constants. 
These high momentum limits are drastically different from the ideal case due to the role of the generalised uncertainty principle.

Moreover, the relativistic adiabatic index $\gamma$ for the degenerate electron gas is obtained as 
\begin{equation} 
\gamma = \frac{\varepsilon+P}{P} \left( \frac{d P}{d\varepsilon} \right)_s= \frac{1}{8}\left(\frac{\tilde{n}^2 }{\tilde{P}} \right)\frac{(1+\alpha^2 \xi^2)^3}{\xi \sqrt{1+\xi^2}}, 
\end{equation}
so that $\gamma \rightarrow \frac{\pi}{16} \alpha^3$ in the limit $\xi\rightarrow \infty$, unlike the ideal case ($\gamma_{\rm ideal}=\frac{4}{3}$).

\section{Mass-radius relation} \label{MR}

We study mass-radius relation of the equilibrium configurations in the framework of general relativity in this section.  For the matter interior to the star, the equilibrium values of the pressure $P(r)$ and energy density $\varepsilon(r)$ are therefore determined by the Tolman-Oppenheimer-Volkoff (TOV) equations \cite{Tolman1939, Oppenheimer1939}
\begin{equation}\label{TOV_1}
\frac{dP}{dr} = - \frac{G}{c^2 r} (\varepsilon + P) \frac{(m + 4\pi P r^3/c^2)}{(r-2Gm/c^2)}
\end{equation}
with
\begin{equation}\label{TOV_2}
\frac{dm}{dr} = \frac{4\pi}{c^2} \varepsilon r^2, 
\end{equation}

It may be observed that the equation of state is in a parametric form where the Fermi momentum $p_F$ of the electron degenerate gas occurs in the expressions for pressure and energy density given by equations (\ref{pressure_1}), (\ref{pressure}), (\ref{energydensity_1}) and (\ref{energydensity}). We express the TOV equations (\ref{TOV_1}) and (\ref{TOV_2}) in terms of the dimensionless quantities $\xi=p_F/m_ec$, $v=m/m_0$ and $\eta=r/r_0$, where $m_0 = (qc^2)^2/G^{3/2}\sqrt{4\pi K}$ and $r_0 = (qc^2)/\sqrt{4\pi G K}$. Thus we obtain 
\begin{equation}\label{dxideta}
\frac{d\xi}{d\eta} = - \frac{1}{\eta}\frac{\sqrt{1+\xi^2}}{\xi}\left(1-q+q\sqrt{1+\xi^2}\right) \frac{v + q  \tilde{P}\eta^3}{\eta-2qv}
\end{equation}
and
\begin{equation}\label{dvdeta}
\frac{dv}{d\eta} = \tilde{\varepsilon}  \eta^2.
\end{equation}

\subsection{Asymptotic solutions} 

For a preliminary idea about the mass-radius relation, we study the asymptotic solutions of the TOV equations in the low and high Fermi momentum limits. 

\subsubsection{Low momentum limit, $\xi\rightarrow0$}\label{3A}

For low values of $\xi$, it can be shown that equations (\ref{dxideta}) and (\ref{dvdeta}) reduce to 
\begin{equation}\label{}
\xi\frac{d\xi}{d\eta} = - \frac{v}{\eta^2}
\end{equation}
and
\begin{equation}\label{}
\frac{dv}{d\eta} = \frac{8}{3} \xi^3  \eta^2.
\end{equation}
which can be combined to form a second order differential equation, given by
\begin{equation}\label{eq:}
\frac{3}{16}\frac{1}{\eta^2}\frac{d}{d\eta}\left(\eta^2 \frac{d\xi^2}{d\eta}\right)+\xi^3=0.
\end{equation}

\begin{figure*} 
\centering
\includegraphics[width=14cm]{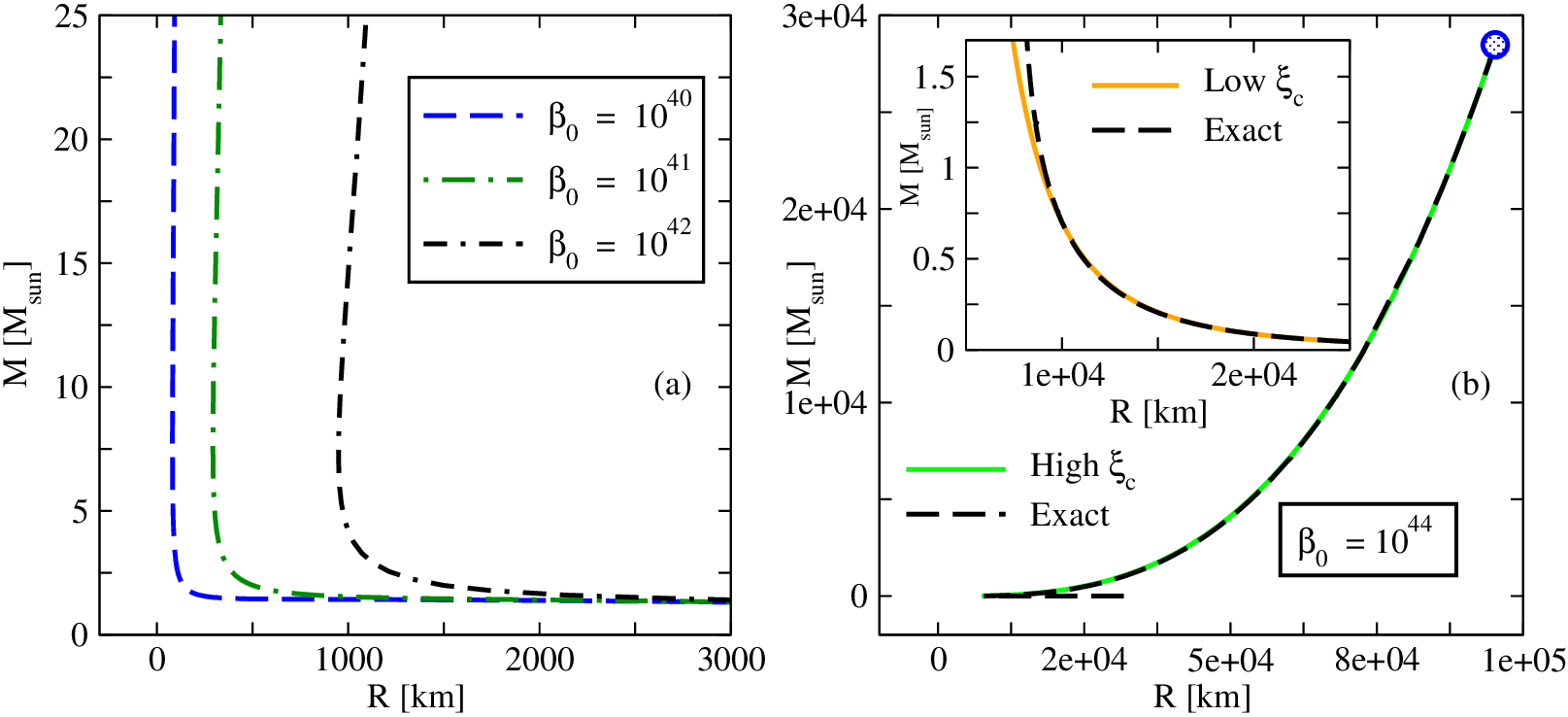}
\caption{(a) Exact mass-radius relations for white dwarfs with GUP equation of state for $\beta_0=10^{42}$, $10^{41}$ and $10^{40}$. (b) Exact mass-radius relations (dashed curves) for $\beta_0=10^{44}$ in comparison with the corresponding analytically obtained asymptotic solution (smooth curve) given by Eqs.~(\ref{RLxi}) and (\ref{MLxi}) in the high $\xi_c$ limit. The open circle represents the maximum values of mass $M_{\rm max}$ and radius $R_{\rm max}$. The lower left region of the exact mass-radius curve is blown up (dashed curve) in the inset where it is compared with the analytically obtained asymptotic solution in the low $\xi_c$ limit (smooth curve). }
\label{Figure_1}
\end{figure*}

Defining $\xi^2(\eta)/\xi^2_c$ as $\theta(\zeta)$, with $\xi_c$ the central dimensionless Fermi momentum, and $\zeta$ is a new dimensionless coordinate $\zeta=\sqrt{16 \xi_c/3}\ \eta$, we reduce the above equation to
\begin{equation}
\frac{1}{\zeta^2} \frac{d}{d\zeta}\left(\zeta^2\frac{d\theta}{d\zeta} \right) +\theta^{3/2}=0,
\end{equation}
which is the Lane-Emden equation of index $3/2$. Numerical solution for this differential equation is given in Weinberg \cite{Weinberg1972}.  For the boundary conditions $\theta(0)=1$ and $\theta'(0)=0$, one can immediately obtain the radius of the white dwarf as
\begin{equation}\label{RLxi}
R = \sqrt{\frac{3}{16\xi_c}}\ r_0\zeta_R
\end{equation}
where $\zeta_R=3.65375$ is the first zero of the Lane-Emden function $\theta(\zeta)$ of index $3/2$. 

Similarly the asymptotic behavior of the mass of the white dwarf can be obtained from the integral expression of Eq.~(\ref{TOV_2}), namely,
\begin{equation}
M=\frac{4\pi}{c^2} \int_0^R \varepsilon(r) r^2 dr = \frac{4\pi}{c^2} \frac{A}{q}  \frac{8}{3} \int_0^R \xi^3  r^2 dr.
\end{equation}
 We rewrite this equation in the new dimensionless variable $\zeta$, yielding 
\begin{equation}
M= \sqrt{\frac{3\xi_c^3}{64}} \  m_0 \int_0^{\zeta_R}\theta^{3/2} \zeta^2 d\zeta,
\end{equation}
thus obtaining the mass of the white dwarf as 
\begin{equation}\label{MLxi}
M = -\sqrt{\frac{3\xi_c^3}{64}} \  m_0 \zeta^2_R \left(\frac{d\theta}{d\zeta}\right)_{\zeta=\zeta_R}
\end{equation}
The value of $\left(-\zeta^2d\theta/d\zeta\right)_{\zeta=\zeta_R}$ is $2.71406$ \cite{Weinberg1972}. 

Thus the above asymptotic analysis predicts that $R\sim\xi_c^{-1/2}$ and $M\sim\xi_c^{3/2}$, giving the mass-radius relation $R\sim M^{-1/3}$, implying that the radius decreases as the mass increases. 

It is important to note that, these expressions of mass and radius are independent of the GUP parameter $\alpha$ (or, equivalently $\beta$). Thus for low mass white dwarfs the GUP has insignificant effect on the mass-radius relation and we expect that the mass-radius curve would coincide with that of Chandrasekhar's for low values of central Fermi momentum $\xi_c$ (or, equivalently low central density $\rho_c$).

\subsubsection{High momentum limit, $\xi\rightarrow\infty$}\label{3B}

For high values of $\xi$, the TOV equations reduce to
\begin{equation}\small\label{}
\frac{d\xi}{d\eta} = - \frac{k_1}{3}\frac{\eta}{1-2q\kappa \eta^2} \bigg( 1-q+q\xi \bigg) \left(1-q+3q\xi  -2q\frac{k_2}{k_1}\right)
\end{equation}
and
\begin{equation}\label{MHxi}
v = \kappa  \eta^3.
\end{equation}

Since typically $\alpha\sim0.1$, the ratio $k_2/k_1\sim(4/\pi \alpha)$, hence the last term in the second bracket can be ignored if $\alpha\xi \gg 8/3\pi$. Since we are looking for the solutions of $\xi\rightarrow\infty$, we shall ignore this term, obtaining 
\begin{equation}\label{}
\frac{d\xi}{d\eta} = - \frac{k_1}{3}\frac{\eta}{1-2q\kappa \eta^2} \left( 1+q\xi \right) \left(1+3q\xi\right),
\end{equation}
where we have used the fact that $q\sim10^{-4}$. The solution of the above equation is given by 
\begin{equation}\label{}
\frac{1+3q\xi}{1+q\xi} = \left(1-2q\kappa \eta^2\right)^{-k_1/6\kappa} + {\rm const}.
\end{equation}
Using the boundary conditions we can immediately obtain the integration constant and hence the radius of the star as
\begin{equation}\label{}
\eta_R = \frac{1}{\sqrt{2q\kappa}}\left\{1-\left(\frac{1+q\xi_c}{1+3q\xi_c}\right)^{6\kappa/k_1}\right\}^{1/2}.
\end{equation}
Since $6\kappa/k_1\approx2$, we have
\begin{equation}\label{}
\eta_R = \frac{1}{\sqrt{2q\kappa}}\left\{1-\left(\frac{1+q\xi_c}{1+3q\xi_c}\right)^2\right\}^{1/2}
\end{equation}

Thus from Eq.~(\ref{MHxi}) the mass become
\begin{equation}\label{}
v_R = \left(\frac{1}{2q}\right)^{3/2}\frac{1}{\sqrt{\kappa}}\left\{1-\left(\frac{1+q\xi_c}{1+3q\xi_c}\right)^2\right\}^{3/2}
\end{equation}

As the the central Fermi momentum approaches larger and larger values, we see that the radius and mass approach maximum values, given by 
\begin{equation}\label{Max}
R_{\rm max} =  \frac{2}{3}\frac{r_0}{\sqrt{q\kappa}}, \hspace{0.3 cm} {\rm and} \hspace{0.3 cm}M_{\rm max} =\frac{8}{27}\frac{m_0}{\sqrt{\kappa} q^{3/2}} 
\end{equation}

\subsection{Exact solutions} 

\begin{figure*}
\centering
\includegraphics[width=15cm]{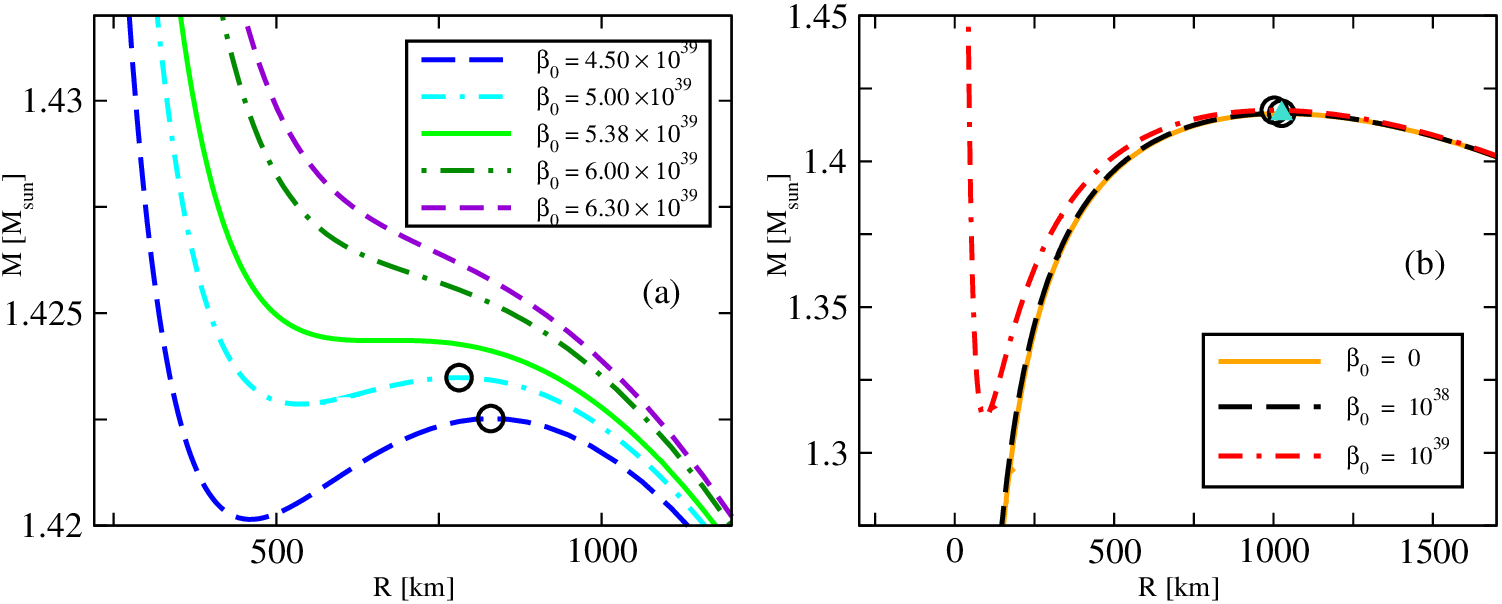}
\caption{(a) Exact mass-radius relations  for $\beta_0$ in the range $4.50\times10^{39}\leq\beta_0\leq6.30\times10^{39}$.  The mass-radius relation for $\beta_0 = \bar{\beta}_0=5.38\times10^{39}$ demarcates these curves into two classes. For $\beta_0 > \bar{\beta}_0$, there occurs no maximal point whereas for $\beta_0 < \bar{\beta}_0$ maximal points ($R^*$, $M^*$) exist (shown by open circles). (b) Exact mass-radius relations for $\beta_0=10^{39}$ and $10^{38}$ in comparison with that of the ideal case, $\beta_0=0$. Proximity of the maximal points ($R^*$, $M^*$) are shown by open circles (for $\beta_0=10^{39}$ and $10^{38}$) with that of the ideal case, shown as a solid triangle ($\beta_0=0$). }
\label{Figure_2}
\end{figure*}

In this section we obtain exact solutions of the TOV equations (\ref{dxideta}) and (\ref{dvdeta}) employing the GUP equation of state expressed by equations (\ref{pressure}) and (\ref{energydensity}) in parametric form. The numerical integrations are carried out with the boundary conditions $\xi(0)=\xi_c$, $v(0)=0$ and $\xi(\eta_R)=0$, where $\eta_R$ denotes the dimensionless radius of the star. The resulting mass-radius relations for different strengths of the dimensionless GUP parameter $\beta_0$ are shown in Figures \ref{Figure_1} and \ref{Figure_2}. 

It is apparent from Figures \ref{Figure_1} and \ref{Figure_2} that, for large values of $\beta_0$, the mass-radius relations given by the GUP equation of state deviate significantly from the ideal case, whereas for smaller values of $\beta_0$, such deviations are smaller. 

In Figures \ref{Figure_1}(a)  and \ref{Figure_1}(b), we display the mass-radius curves for higher magnitudes of the GUP parameter such as $\beta_0 = 10^{44}$, $10^{42}$, $10^{41}$ and $10^{40}$. We see that the mass-radius curves coincide with the Chandrasekhar's curve only for low values of the central Fermi momentum $\xi_c$, as shown in the right-hand part of the inset in Figure \ref{Figure_1}(b). This is evident from the fact that the TOV equation reduces to Newtonian equation in the low density regime. Moreover, we see from the right-hand part of Figure \ref{Figure_1}(a) that all curves nearly coincide irrespective of the strength of the GUP parameter $\beta_0$. This is due to the fact that $\beta_0$ disappears from the asymptotic equations in this regime as we have seen earlier in Section \ref{3A}.

For higher $\xi_c$ values, the exact mass-radius curve reaches a point where the radius is minimum $R_{\rm min}$. The $R_{\rm min}$  value is smaller for smaller $\beta_0$ values as seen in Figure \ref{Figure_1}(a). On further increasing $\xi_c$, both the mass and radius increase reaching terminal values as shown in Figure \ref{Figure_1}(b) denoted by a open circle. In this regime, we see that analytically obtained high momentum solution (Section \ref{3B}) coincides with the exact mass-radius curve as shown in  Figure \ref{Figure_1}(b). Moreover, the terminal values of radius  $R_{\rm max}$ and mass  $M_{\rm max}$ given by equation (\ref{Max}) are found to be nearly the same as given by the exact solutions. However these terminal values are excessively high, as evident from Figure \ref{Figure_1}(b).

Exact mass-radius curves for intermediate strengths of the GUP parameter $\beta_0$ (in the range $4.50\times10^{39}\leq\beta_0\leq6.30\times10^{39}$) are shown in Figure \ref{Figure_2}(a). We see a cross-over in the behavior of  the curves around the value $\beta_0 = \bar{\beta}_0=5.38\times10^{39}$. For $\beta_0>\bar{\beta}_0$, the mass-radius curves do not have a maximal point, whereas for $\beta_0<\bar{\beta}_0$, there exist maximal points. Figure \ref{Figure_2}(b) compares the mass-radius relation for smaller values of $\beta_0$ ($=10^{39}$ and $10^{38}$) with the ideal case ($\beta_0 = 0$). We see that the maxima of the mass-radius curves for these values of $\beta_0$ nearly coincide with the maxima of the ideal case.  It is also important to note that the maxima shifts slightly towards the right [Figure \ref{Figure_2}(a)] as the value of $\beta_0$ is decreased until the maxima coincide with the ideal value [Figure \ref{Figure_2}(b)].

A more rigorous treatment is required to assert whether these maxima correspond to the onset of gravitational instability. Moreover, it is critical to analyze the role of GUP parameter in the determination of the maximum possible mass of white dwarfs. In the following section we perform a rigorous stability analysis of the equilibrium configurations by investigation the dynamical instability in the framework of general relativity. It consists of studying dynamics of time dependent infinitesimal radial perturbation about the equilibrium configuration at every point inside the star in a homologous manner \cite{Chandrasekhar1964a}. The time evolution of these perturbations determined by the central Fermi momentum $\xi_c$ and the GUP parameter $\beta_0$ establish whether the system is stable or otherwise.

\section{Dynamical stability analysis}\label{Stability}

As we have already noted, dynamical stability analysis consists of the investigation of the time evolution of homologous infinitesimal perturbation about the equilibrium configuration \cite{Chandrasekhar1964a,Chandrasekhar1964b,Chandrasekhar1964c}. The corresponding metric interior to the star is expressed as
\begin{equation}\label{metric}
ds^2 = e^{\nu+\delta \nu} c^2 dt^2 - e^{\mu +\delta \mu}dr^2 -r^2(d\theta^2 + \sin^2 \theta \ d\phi^2),
\end{equation}
where $\nu(r)$ and $\mu(r)$ are the equilibrium metric potentials and the perturbations  $\delta \nu(r,t)$ and $\delta \mu(r,t)$ are due to small radial Lagrangian displacements $\zeta(r,t)$. This induces perturbations $\delta P(r,t)$ and $\delta \varepsilon (r,t)$ to the equilibrium pressure $P(r)$ and energy density $\varepsilon(r)$. The smallness of the perturbation allows one to consider sinusoidal displacements $\zeta(r,t) = r^{-2} e^{\nu/2} \psi(r) e^{i\omega t}$. The corresponding equation for the radial oscillation can be obtained in the Strum-Liouville form \cite{Bardeen1966}
\begin{equation}\label{main}
\frac{d}{dr}\left(U\frac{d\psi}{dr}\right) + \left(V + \frac{\omega^2}{c^2} W\right)\psi = 0,
\end{equation}
satisfying the boundary conditions $\psi = 0$  at $r=0$  and the Lagrangian change in pressure $\delta P =-e^{\nu/2}\frac{\gamma P}{r^2}\frac{d\psi}{dr}= 0$ at  $r=R$. In the above equation, 
\begin{eqnarray}
U(r) = \ &&e^{(\mu+3\nu)/2} \frac{\gamma P}{r^2}, \label{coefficient1} \\
V(r) = \ &&-4 \frac{e^{(\mu+3\nu)/2}}{r^3}\frac{dP}{dr} - \frac{8\pi G}{c^4} \frac{e^{3(\mu+\nu)/2}}{r^2}P(P+\varepsilon) \label{coefficient2} \nonumber \\
&&+ \frac{e^{(\mu+3\nu)/2}}{r^2} \frac{1}{P+\varepsilon} \left(\frac{dP}{dr}\right)^2,\\
W(r) = \ &&\frac{e^{(3\mu+\nu)/2}}{r^2}(P+\varepsilon), \label{coefficient3}
\end{eqnarray}
with the adiabatic index $\gamma$, given by 
\begin{equation} 
\gamma = \frac{\varepsilon+P}{P} \left( \frac{d P}{d\varepsilon} \right)_s. 
\end{equation}

Integrating Eq.~(\ref{main}) upon left-multiplying by $\psi$, one obtains the integral 
\begin{equation}\label{Rayleighritz}
J[\psi] = \int_0^R \left\{ U \psi'^2 -V\psi^2 -  \frac{\omega^2}{c^2} W \psi^2 \right\} dr
\end{equation}
where $\psi'=d\psi/dr$, and the boundary conditions eliminate the surface term. It can be shown that Eq.~(\ref{main}) is reproduced from the variational principle $\delta J[\psi]=0$. Thus the lowest characteristic eigenfrequency of the normal mode is obtained from 
\begin{equation}\label{Rayleighritz}
\frac{\omega_0^2}{c^2} =  \min_{\psi(r)} \frac{ \int_0^R \left\{ U \psi'^2 -V\psi^2 \right\} dr}{\int_0^R W \psi^2  dr}. 
\end{equation}

The star remains in stable equilibrium so long as this equation yields positive values of $\omega_0^2$. On the other hand, a negative $\omega_0^2$ signifies unstable equilibrium. A power series solution of Eq.~(\ref{main}) about $r=0$ gives $\psi(r)\propto r^3$ in the leading order for which $\zeta(r)$ and $\zeta'(r)$ are finite.  A good approximation for the trial function of the fundamental mode can be taken as the simple form $\psi(r) = c_0 r^3$ \cite{Chandrasekhar1964a,Chandrasekhar1964b,Wheeler1968}. With this choice, the onset of instability, hence the critical density $\rho_c^*$ for gravitational collapse, can be identified with a zero eigenfrequency solution of Eq.~(\ref{Rayleighritz}). 

For the matter interior to the star, the equilibrium values of the pressure $P(r)$ and energy density $\varepsilon(r)$ are determined by the Tolman-Oppenheimer-Volkoff (TOV) equations (\ref{TOV_1}) and (\ref{TOV_2}) and the interior Schwarzschild  metric potentials satisfying the  Einstein's field equations are given \cite{Tolman1939, Oppenheimer1939} by
\begin{equation}
e^{-\mu(r)} = 1-\frac{2Gm}{c^2 r}
\end{equation}
and 
\begin{equation}
e^{\nu(r)} = \left(1-\frac{2GM}{c^2 R}\right)\exp\left[-2\int_0^{P(r)} \frac{dP}{\varepsilon+P}\right]. 
\end{equation}

\begin{figure} 
\centering
\includegraphics[width=8.5cm]{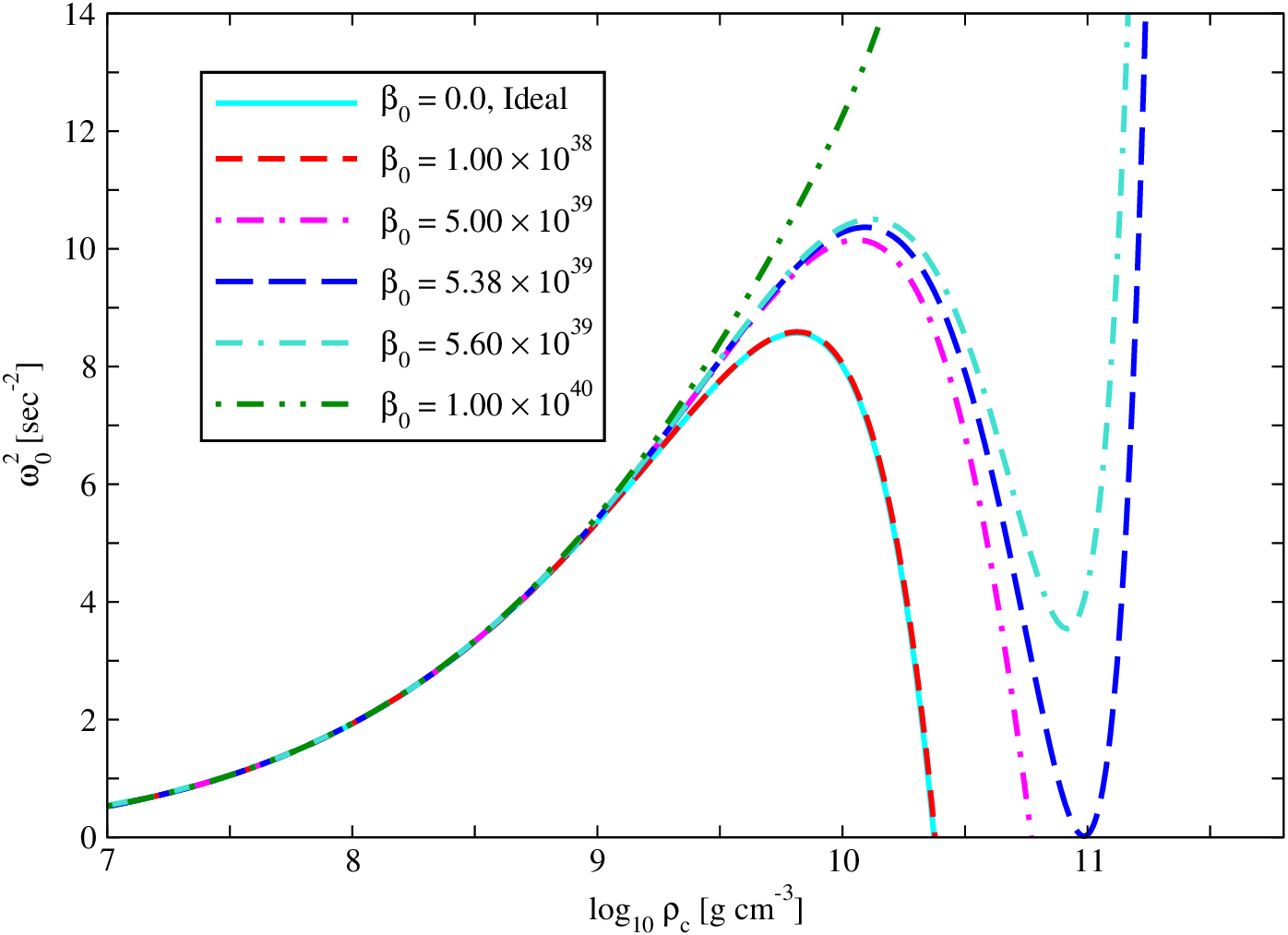}
\caption{Eigenfrequency of the fundamental mode $\omega_0^2$ against central density $\rho_c$ for various values of the GUP parameter $\beta_0$.}
\label{Figure_3}
\end{figure}

\subsection{Eigenfrequency of the fundamental mode}

The interior Schwarzschild  metric potentials can be written in dimensionless varibles as 
\begin{equation}\label{lambda}
e^{-\mu(\eta)} = 1-2q\frac{v}{\eta}
\end{equation}
and 
\begin{equation}\label{nu2}
e^{\nu(\eta)} =\left(1-2q \frac{v_R}{\eta_R}\right) \left(\frac{1}{1-q+q\sqrt{1+\xi^2}}\right)^2,
\end{equation}
where the expression for $e^{\nu}$ is obtained from the equation of state given by equations (\ref{pressure}) and (\ref{energydensity}).

The solution of the TOV equations ~(\ref{dxideta}) and  (\ref{dvdeta}), and equations (\ref{lambda}) and (\ref{nu2}) give all quantities necessary for the evaluation of the functions $U(r)$, $V(r)$ and $W(r)$ in equations (\ref{coefficient1}\---\ref{coefficient3}). We may rewrite Eq.~(\ref{Rayleighritz}) in dimensionless form as 
\begin{equation}\label{omega}
\omega_0^2 = \left(\frac{q c^2}{r_0^2}\right)  \frac{ \mathcal{I} +  \mathcal{J}}{\mathcal{K}},
\end{equation}
where 
\begin{equation}\label{I}
\mathcal{I}=\int_0^{\eta_R} e^{(\mu+3\nu)/2} \frac{\gamma \tilde{P}}{\eta^2} \psi'^2  d\eta,
\end{equation}
\begin{eqnarray}\label{J}
\mathcal{J}=\int_0^{\eta_R}\frac{e^{(\mu+3\nu)/2}}{\eta^2}\left[\frac{4}{\eta} \frac{d\tilde{P}}{d\eta} + 2q e^\mu \tilde{P}(\tilde{\varepsilon}+q\tilde{P}) \right. \nonumber\\
\left. -\frac{q}{\tilde{\varepsilon}+q\tilde{P}} \left(\frac{d\tilde{P}}{d\eta}\right)^2\right] \psi^2 d\eta,  
\end{eqnarray}
and
\begin{equation}\label{K}
\mathcal{K}=\int_0^{\eta_R} e^{(3\mu+\nu)/2} \frac{\tilde{\varepsilon}+q\tilde{P}}{\eta^2} \psi^2 d\eta.
\end{equation}

We thus numerically evaluate the integrals in Eq.~(\ref{I})---(\ref{K}) with the trial function $\psi= c_0 \eta^3$, where $c_0$ a disposable constant, for different choices of the GUP parameter $\beta_0$. Consequently we obtain the  eigenfrequency of the fundamental mode  $\omega_0^2$ from equation (\ref{omega}). As stated earlier, stable configurations correspond to positive values of $\omega_0^2$ whereas a zero frequency solution indicates the onset of a dynamical instability signifying the onset of a gravitational collapse.

We display the results of the numerical integrations in Figure \ref{Figure_3}, where the eigenfrequency $\omega_0^2$ is plotted with respect to the central density $\rho_c$ (= $\varepsilon_c/c^2$) for different values of $\beta_0$. We observe that for low mass white dwarfs with central densities $\rho_c\lesssim10^{9}$ g cm$^{-3}$, the pulsation frequencies overlap signifying the irrelevance of the effect of GUP in this range of central densities. The pulsation frequencies start to deviate from each other in the higher density regime depending on the value of $\beta_0$. 

For $\beta_0 \leq \bar{\beta}_0=5.38\times10^{39}$, there exist zero eigenfrequency solutions at central densities $\rho_c^*$, suggesting the onset of gravitational collapse. The existence of imaginary eigenfrequency solution corresponding to unstable configuration is possible only for $\beta_0 < \bar{\beta}_0$. For  $\beta_0 > \bar{\beta}_0$, zero eigenfrequency solutions are not possible even for arbitrarily high central densities $\rho_c$, signifying stability of excessively massive white dwarfs. We also see that the curve for $\beta_0=10^{38}$ nearly coincides with that for the ideal case ($\beta_0=0$). This means that all curves in the range $0\leq\beta_0\leq10^{38}$ overlap (to a good approximation) giving rise to approximately the same onset density  $\rho_c^*$ for gravitational collapse. A legitimate upper bound is given by the electroweak limit $\beta_0\sim10^{34}$ \cite{Das2008} which is well within the range $0\leq\beta_0\leq10^{38}$. Since this onset density is nearly $2.3588\times10^{10}$ g cm$^{-3}$, Chandrasekhar's general relativistic mass $\sim1.42$ M$_\odot$ is easily recovered in this range which extends four orders of magnitude beyond the electroweak bound.

\begin{table}
\caption{Critical values of the central density $\rho_c^*$, mass $M^*$, and radius $R^*$ for different values of the GUP parameter $\beta_0$ at the onset of dynamical instability determined by the vanishing eigenfrequency of the fundamental mode.}
\vspace{0.2cm}
\begin{tabular}{cccc}
\hline
 $\beta_0$ & $\rho_c^*$ (g cm$^{-3}$)   &   $M^*$ (M$_{\odot}$)  &     $R^*$ (km)         \\
\hline
$5.38\times10^{39}$ 	&     $1.0105\times10^{11}$    &     $1.4244$      &      $655.5629$      \\ 
$5.00\times10^{39}$		&     $5.8618\times10^{10}$    &     $1.4235$      &      $776.3669$      \\ 
$1.00\times10^{38}$ 	&     $2.3801\times10^{10}$    &     $1.4165$      &      $1021.6162$    \\ 
$1.00\times10^{34}$ 	&     $2.3588\times10^{10}$    &     $1.4164$      &      $1024.3821$     \\ 
\hline
\end{tabular}
\label{Table_1} 
\end{table}

The above discussions lead to parallel observations from Figure \ref{Figure_2} (a) where $\beta_0 = \bar{\beta}_0$ demarcates a change in behavior of the mass-radius curves. The non-existence of a maximal point in the mass-radius curve for $\beta_0 > \bar{\beta}_0$ is evident from the fact that there exists no critical density $\rho_c^*$ corresponding to a zero eigenfrequency solution. On the other hand, for  $\beta_0 < \bar{\beta}_0$, the existence of maximal points ($R^*$, $M^*$) in the mass-radius curves are consequences of zero eigenfrequency solutions at $\rho_c^*$. The branches towards the right of the maximal point ($R^*$, $M^*$) correspond to lower central densities $\rho_c<\rho_c^*$ and thus the stability of this branch is confirmed by the fact that  $\omega_0^2$ is positive, as shown in Figure \ref{Figure_3}. On the other hand, the branches towards the left of the maximal point ($R^*$, $M^*$) correspond to instability as $\omega_0^2$ becomes negative (not shown in Figure \ref{Figure_3}) and they correspond to $\rho_c>\rho_c^*$.

As $\beta_0$ is deceased towards $10^{38}$, the maximal points ($R^*$, $M^*$) approach closer to each other and nearly coincide at $\beta_0=10^{38}$. The corresponding critical values are displayed in Table \ref{Table_1} where it is evident that the critical mass approaches the limit $1.416$ M$_\odot$ and the radius 1024 km. 

Thus in addition to asserting the existence of the Chandrasekhar limit, the stability analysis confirms the fact that the radius decreases as the mass increases for stable configurations of white dwarfs.

\section{Conclusion}\label{Conclusion} 

As we have already discussed in detail, there have been a few attempts to restore the Chandrasekhar limit when white dwarfs are described by GUP-enhanced equation of state. However all these attempts are found to be inadequate from different perspectives, and some of them led to unphysical mass-radius relations. It thus becomes essential to resolve this issue in a cogent fashion so that all assumptions in the theory are physically plausible. In fact, we have shown that this issue is resolved in a rigorous manner by adopting general relativity vis-\`a-vis GUP-enhanced equation of state with {\it positive} GUP parameter $\beta$. Importantly, we find that the Chandrasekhar mass is assured for $\beta_0$ values below $\bar{\beta}_0$, for which onset of gravitational collapse is possible. We also note that the electroweak upper bound for $\beta_{0}$ is much below $\bar{\beta}_0$ so that physical existence of Chandrasekhar limit is guaranteed. 

The above conclusion stems from a rigorous stability analysis of the equilibrium configurations as displayed in Figure \ref{Figure_3}, where  the eigenfrequency of the fundamental mode $\omega_0^2$ is plotted with respect to the central density $\rho_c$ for different values of $\beta_0$. We see that a vanishing eigenfrequency exists when $\beta_0\leq5.38\times10^{39}=\bar{\beta}_{0}$, giving rise to a dynamical instability at critical central densities $\rho_c^*$. However, for $\beta_0>\bar{\beta}_0$, no dynamical instability occurs because of the nonexistence of a zero eigenfrequency solution, implying that these configurations remain stable for arbitrarily high values of $\rho_c$ leading to excessively massive white dwarfs. However these solutions are physically unacceptable because the corresponding $\beta_0$ values are well above the electroweak bound.  

An important point to observe from Figure \ref{Figure_3} is that the eigenfrequencies for $\beta_0=10^{38}$ practically coincides with that of the ideal case, $\beta_0=0$. Thus in the range $0<\beta_0<10^{38}$, the critical density $\rho_c^*$ for the onset of gravitational collapse (determined by the vanishing eigenfrequency) remains practically unaltered. We find $\rho_c^*=2.3801\times10^{10}$ g cm$^{-3}$ for $\beta_0=10^{38}$, which is nearly the same as Chandrasekhar's critical value of $2.3\times 10^{10}$ g cm$^{-3}$ (for $\beta_0=0$). It is thus evident that Chandrasekhar's general relativistic critical mass of $1.42$ M$_\odot$ \cite{Chandrasekhar1964c} remains practically unaffected.

In the context of the stability analysis, we can analyze the mass-radius curve obtained in Section \ref{MR}. Since the Chandrasekhar limit exists only below $\bar{\beta}_0=5.38\times10^{39}$, all mass-radius plots in Figure \ref{Figure_1} above this value would not correspond to reality. This is also evident from the fact that $\bar{\beta}_{0}$ is much higher than the electroweak bound $\beta_0\sim10^{34}$. For $\beta_0<\bar{\beta}_0$, the mass-radius curves develop maximal points [see Figure \ref{Figure_2} (a)] at which the eigenfrequencies $\omega_0^2$ vanish as shown later in Section \ref{Stability}. These maximal points correspond to limiting Chandrasekhar mass lying below $\sim1.425$ M$_\odot$. It is important to note that {\it the radius decreases as the mass increases} in the part of a mass-radius curve towards the right of the maximal point that corresponds to the {\it stable} branch. The mass-radius behavior in the stable branches is consistent with several astronomical observations of white dwarfs \cite{Kilic2007,Vennes1997,Vennes1999,Marsh1997,Kepler2007,Shipman1972,Shipman1977,Shipman1979}.  In this context it may be recalled that the other attempts, based on negative $\beta$ \cite{Ong2018b} and $\Lambda$-modified GUP \cite{Ong2018a}, led to mass-radius relations that contradict the astronomical observations. Moreover, our stability analysis suggests that upon reaching beyond the Chandrasekhar mass the star would collapse to form highly dense compact objects such as neutron star or black hole. Formation of such compact objects are impossible in the other scenarios (based on negative $\beta$ \cite{Ong2018b} and $\Lambda$-modified GUP \cite{Ong2018a}) where the density would be infinitely diluted as the Chandrasekhar mass is approached. 

The present scenario of describing white dwarfs in terms of general relativity and GUP-enhanced equation of state with a positive GUP parameter $\beta$ rigorously leads to the existence of Chandrasekhar mass as well as the correct behavior of the mass-radius relation consistent with astronomical observations. Moreover it suggests the onset of gravitational collapse beyond the Chandrasekhar mass leading to the possibility of highly dense astrophysical objects such as neutron stars or black holes.

\section*{Acknowledgement}
Arun Mathew would like to thank the Indian Institute of Technology Guwahati for financial assistance through a Research and Development fund.

\end{document}